\documentclass[aps,pre,twocolumn,showpacs,superscriptaddress,groupedaddress]{revtex4-2}
\usepackage{amsmath}
\usepackage{amsfonts}
\usepackage{amssymb}
\usepackage{graphicx}
\usepackage{dcolumn}
\usepackage{sidecap}
\usepackage{parskip}
\usepackage{xcolor}
\usepackage{hyperref}
\usepackage{hhline}
\usepackage{mathtools}
\usepackage{multirow}
\usepackage{verbatim}
\usepackage{rotating}
\usepackage{setspace}
\usepackage{epsfig}
\usepackage{epstopdf}
\usepackage{caption}
\usepackage{subcaption}

\usepackage[normalem]{ulem}

\usepackage{soul}
\usepackage{bm}

\makeatletter
\def\@eqnnum{{\normalsize \normalcolor (\theequation)}}
 \makeatother

\begin{document}
\title{Prolonged hysteresis in the Kuramoto model with inertia and higher-order interactions}

\author{Narayan G Sabhahit$^{1}$}
\email{narayan.g.sabhahit@gmail.com}
\author{Akanksha S. Khurd$^{2}$}
\author{Sarika Jalan$^{3}$}
\email{sarika@iiti.ac.in}

\affiliation{1. Department of Physical Sciences, Indian Institute of Science Education and Research, Kolkata}
\affiliation{2. Department of Physics, Indian Institute of Science Education and Research, Tirupati}
\affiliation{3.Complex Systems Lab, Department of Physics, Indian Institute of Technology Indore, Khandwa Road, Simrol, Indore-453552, India}

\date{\today}

\begin{abstract}
The inclusion of inertia in the Kuramoto model has been long reported to change the nature of phase transition, providing a fertile ground to model the dynamical behaviors of interacting units. More recently, higher-order interactions have been realized as essential for the functioning of real-world complex systems ranging from the brain to disease spreading. Yet, analytical insights to decipher the role of inertia with higher-order interactions remain challenging. Here, we study Kuramoto model with inertia on simplicial complexes, merging two research domains. We develop an analytical framework in a mean-field setting using self-consistent equations to describe the steady-state behavior, which reveals a prolonged hysteresis in the synchronization profile. Inertia and triadic interaction strength exhibit isolated influence on system dynamics by predominantly governing, respectively, the forward and backward transition points. This work sets a paradigm to deepen our understanding of real-world complex systems such as power grids modeled as the Kuramoto model with inertia.
 
\end{abstract}

\maketitle
\section{Introduction}
\label{section 1}
The emergence of collective behavior in complex real-world systems has been a long-standing research interest \cite{boccaletti2006complex}. It was initially in the landmark paper \cite{kuramoto1975self} that Kuramoto modeled the phenomenon of synchronization using a system of network-coupled oscillators in an analytically tractable setting, illustrating that the system underwent a second-order phase transition from incoherent to a coherent state. Since then, numerous works on various extensions of the Kuramoto Model have been done, revealing several phenomena \cite{childs2008stability,sethia2008clustered,omel2012nonuniversal,olmi2015chimera,rodrigues2016the}. Of particular interest to us is the Kuramoto Model with inertia (also known as the second-order Kuramoto model). Inspired by the modeling of synchronized flashing in \textit{Pteroptix malaccae} by \textit{Ermentrout} \cite{ermentrout1991adaptive}, a second-order extension of the Kuramoto model was first proposed by \textit{Tanaka et al.} \cite{tanaka1997first,tanaka1997self}. They showed that the system experienced a first-order phase transition upon introducing inertia rather than the smooth second-order phase transition observed in the Kuramoto model. They put forth a self-consistent method akin to the one proposed by Kuramoto to study the steady-state behavior of the coupled oscillator system. Since then, the second-order Kuramoto model has been extensively explored in diluted networks \cite{ji2013cluster} and various real-world complex systems like Josephson junctions \cite{trees2005sync} and power grids \cite{ji2014basin, grzybowski2016on, rohden2012self, dorfler2013synchronization, witt2022col}. In \cite{filatrella2008analysis}, \textit{Filatrella et al.} explained how the second-order Kuramoto model originates in power grids by simply accounting for power conservation at each node of the grid, rendering it more than just a mathematical convenience.  

However, all these results were obtained by focusing on the interactions to be purely dyadic in nature. Recent research highlights that such a reductionist view might not reveal the complete picture of the underlying mechanism of exotic phenomena observed in some real-world complex systems where the interactions between agents are inherently higher-order in nature \cite{benson2018simplicial,Jalan_phyRep2023,battiston2020networks,SJ_Priyanka2023}. Using the Ott-Antonsen (OA) dimensionality reduction method \cite{Ott_Antonsen}  \textit{Skardal \& Arenas} \cite{skardal2020higher} showed that incorporating higher-order interactions into the Kuramoto model resulted in abrupt (de)synchronization transitions. It was remarkable to observe that adding an inertia term to the Kuramoto model or incorporating higher-order interactions independently gave rise to a first-order phase transition in the system.
 Here, we are interested in understanding how the interplay of inertia and higher-order interactions manifests itself in the system and affects the synchronization profile, which has not been explored before. 

\begin{figure*}[t]
    \centering 
    \includegraphics[width= 2\columnwidth]{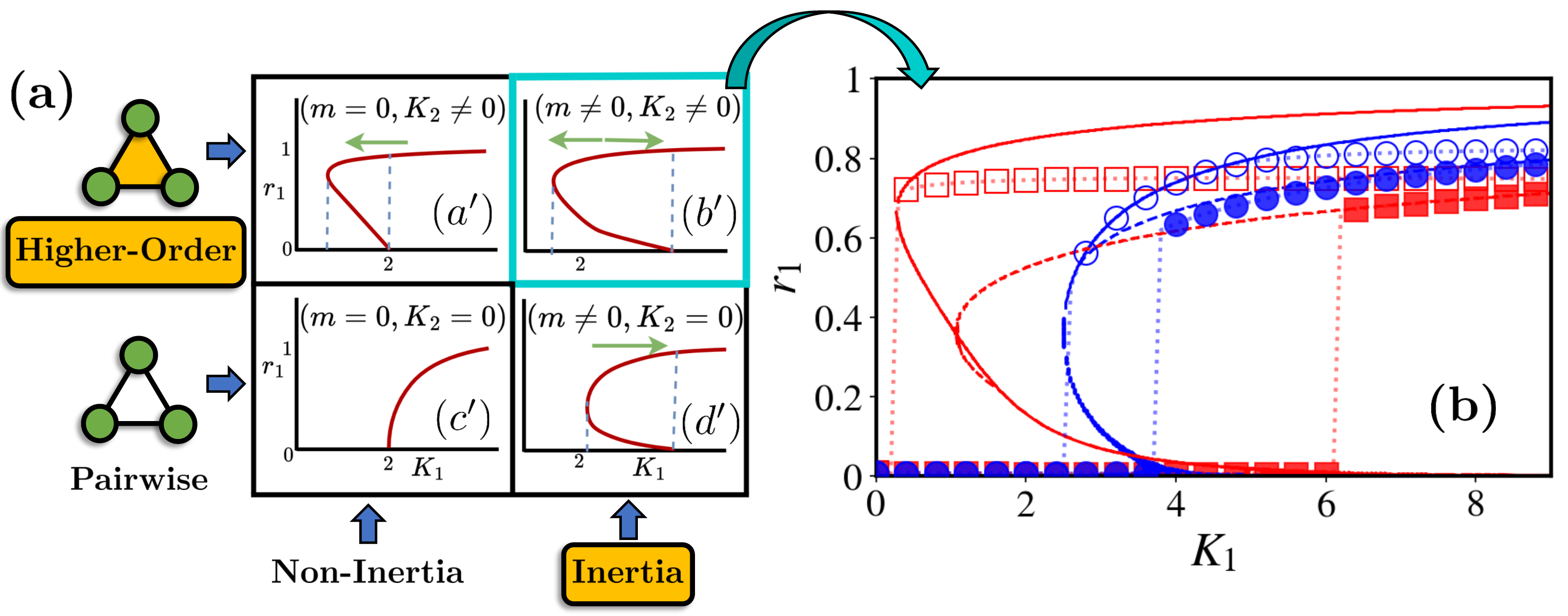}
    \caption{(Color online) Prolonged Hysteresis. a) Schematic depiction of emerging collective behavior in the Kuramoto Model (KM). ($a^{\prime}$), ($c^\prime$), and ($d^\prime$) plot the usual behavior of KM \cite{kuramoto1975self} in the sole impression of higher-order \cite{skardal2020higher} or inertia \cite{tanaka1997first}, whereas  
    ($b^{\prime}$) illustrates a simultaneous forward and backward shift in the transition points upon introduction of $m$ and $K_{2}$ in KM (Eq.~\ref{Gen_KM_hoi}), revealing a prolonged hysteresis. 
    The green arrow indicates the direction of the shift in the transition points. b) $r_1$ versus $K_1$ plot for $K_2 = 1$ and $m = 1$ (blue-circles) and $K_2 = 7$ and $m = 3$ (red-squares). Filled circles and squares represent the simulation results for the forward, and hollow circles and squares represent the backward processes. The dashed and continuous curves represent the forward and backward analytical predictions, respectively.}
    \label{fig: figure1}
\end{figure*}

 This article unifies these two disparate fields by providing a generalized analytical framework to predict the steady-state behavior of coupled oscillator systems with inertia interacting via higher-order interactions. The first challenge lies in the fact that using dimensionality reduction methods like the OA  does not carry over easily towards analyzing second-order Kuramoto models \cite{PengJi_SciRep2014} as the density function, which is usually Fourier expanded in terms of the phase now also depends on the velocity of the oscillators. Hence, in this study, we develop novel analytics using the self-consistency methodology originally proposed by Kuramoto and later extended to the second-order Kuramoto model by \textit{Tanaka et al.} \cite{tanaka1997first}. We show that the effect of inertia and higher-order interactions manifest independently, resulting in a prolonged hysteresis-driven synchronization transition within the system. The forward and backward transition points are found to be predominantly dependent on inertia and higher-order interaction, respectively Fig.~\ref{fig: figure1}.

\section{Model}
\label{section 2}

We study the inertial effects in a globally coupled oscillator network considering the simultaneous presence of dyadic and triadic interactions. Phases of $N$-coupled oscillators, each with mass $m$, evolve based on the following coupled nonlinear equations,
\begin{equation}\label{Gen_KM_hoi}
    \begin{split}
    m\ddot{\theta_i} =  -\dot{\theta_i} + \omega_i + \frac{K_{1}}{N}\sum_{j = 1}^{N}\sin(\theta_{j} - \theta_{i}) \\
    +\frac{K_{2}}{N^{2}} \sum_{j = 1}^{N}\sum_{k = 1}^{N}\sin(2\theta_{j} - \theta_{k}- \theta_{i}).
    \end{split}
\end{equation}
In Eq.~\ref{Gen_KM_hoi}, $\theta_{i}$ and $\dot\theta_{i}$ refer to the  instantaneous phase and angular velocity of  the $i^{th}$ oscillator, respectively. $\omega_{i}$ is the intrinsic frequency of the $i^{th}$ oscillator derived from a unimodal symmetric probability distribution $g(\omega)$ with mean $\Omega$.  The coupling constants $K_1 \geq 0$ and $K_2$ are the dyadic and triadic coupling strengths, respectively. The form of the higher order coupling term directly falls from the phase reduction of mean-field complex Ginzburg-Landau equation \cite{ginzberglandau}.

We decouple the differential equations in Eq.~\ref{Gen_KM_hoi} and write them in terms of mean-field quantities by introducing the following general order parameter for $p \in \{1,2\}$,
\begin{equation}\label{Order Parameter}
     r_{p}e^{i\psi_{p}} = \frac{1}{N}\sum_{j = 1}^{N}e^{ip\theta_{j}}.
\end{equation}
From the above definition, $r_{1}$ measures the global phase coherence and can be interpreted as the centroid of phases of oscillators on a unit circle in the complex plane, and $\psi_{1}$ measures the average phase of the oscillators. $r_{2}$, referred to as the Daido order parameter \cite{daido1996multi} captures cluster synchronization. As we are interested in the steady state behavior of the system, we omit the time dependence in the definition of the general order parameter. In the incoherent state, the phases of the oscillators are scattered uniformly on the unit circle and hence $r_{1} \approx r_{2} \approx 0$. Meanwhile, in the coherent state, a single group of oscillators is formed locked to the mean phase $\psi_{1}$ rotating uniformly at angular velocity $\Omega$, hence $r_{1} \approx r_{2} \approx 1$. Using Eq.~\ref{Order Parameter}, Eq.~\ref{Gen_KM_hoi} can be written in terms of mean-field quantities as,
\begin{equation}\label{mean field equation}
    \begin{split}
        m\ddot{\theta}_{i} =  -\dot{\theta}_{i} + \omega_i + K_{1}r_{1}\sin(\psi_{1} - \theta_{i}) \\  +K_{2}r_{1}r_{2}\sin(\psi_{2} - \psi_{1}- \theta_{i}).
    \end{split}
\end{equation}
Because of the rotational symmetry in the model, the mean of the $g(\omega)$ distribution can be set to zero by moving into the rotating frame at the frequency $\Omega$. This can be facilitated by making the transformation $\theta_{i} \rightarrow \theta_{i} + \Omega t$ in Eq.~\ref{Gen_KM_hoi}. Once in the rotating frame,  $\psi_{1}$ and $\psi_{2}$ can be set to zero by appropriately shifting the origin, i.e., $\theta_i(0) \rightarrow \theta_i(0) + \psi_1(0)$. Eq.~\ref{mean field equation}, now takes the following form,
\begin{equation}\label{mean field equation 2}
    m\ddot{\theta}_{i} =  -\dot{\theta}_{i} + \omega_i - q\sin(\theta_{i}),
\end{equation}
where, for the ease of notation, $q= r_{1}(K_{1} +K_{2}r_{2})$.

\section{Analytical Results}
\label{section 3}
Note that for fixed parameter values $K_{1}$ and $K_{2}$, Eq.~\ref{mean field equation 2} has two variables $r_{1}$ and $r_{2}$. Hence, to chalk out the steady state behavior of Eq.~\ref{mean field equation 2}, we develop a system of self-consistent equations and seek the values of $(K_{1},r_{1},r_{2})$ which simultaneously satisfy them. We start by taking the thermodynamic limit ($N\rightarrow\infty$); the coupled oscillator system in the steady state is then described by a probability density $\rho(\theta,\omega)$ where for a given intrinsic frequency $\omega$, $\rho(\theta,\omega)d\theta$ represents the fraction of oscillators with their phase between $\theta$ and $\theta + d\theta$. The general order parameter in Eq~\ref{Order Parameter} takes the  following form in the continuum limit,
\begin{equation}\label{Order Parameter Thermodynamic}
     r_{p}e^{i\psi_{p}} = \int_{-\infty}^{\infty}\int_{-\pi}^{\pi} e^{ip\theta}\rho(\theta,\omega)g(\omega)d\omega d\theta .
\nonumber \end{equation}

In the steady state, the oscillator population splits up into two groups depending on their intrinsic frequency. One group of oscillators is locked to the mean phase; meanwhile, the other oscillators drift over the locked oscillators. Hence the overall phase coherence ($r_{p}$) can be split into contributions from the locked ($r_{p}^{l}$) and drifting ($r_{p}^{d}$) oscillators, i.e, $r_{p} = r_{p}^{l} + r_{p}^{d}$.
Before calculating $r_p^l$ and $r_p^d$, we point out that systems whose motion is governed by Eq.~\ref{mean field equation 2} are known to depict hysteresis and have been well studied in \cite{tanaka1997first,tanaka1997self,strogatz2018nonlinear}. For completeness, we briefly summarise the reason for the hysteretic behavior here. Dropping the subscript $i$ and by introducing a new timescale $\tau = \sqrt{\frac{q}{m}}t$, Eq.~\ref{mean field equation 2} is transformed to a second order differential equation with just two parameters as,

\begin{equation}\label{dimensionless mean field equation}
    \ddot{\theta} =  -\alpha \dot{\theta} + \beta -\sin(\theta),
\end{equation}

where $\alpha = \frac{1}{\sqrt{qm}}$ and $\beta = \frac{\omega}{q}$. This equation has two fixed point solutions, a saddle point and a stable node for $\beta<1$, obtained by setting $\dot\theta = 0$ and $\ddot\theta = 0$. At $\beta = 1$, the system undergoes a saddle-node bifurcation, annihilating the two fixed point solutions and admitting a unique stable limit cycle solution for all $\beta >1$ \cite{levi1978dynamics}. However, it so happens that as we decrease the value of $\beta$ to be less than one, the limit cycle persists for some small values of $\alpha$. Hence, bistability exists in the system, where a stable limit cycle and a stable node coexist. A further decrease in $\beta$ will disintegrate the limit cycle via a homoclinic bifurcation. Fig.~\ref{fig : figure2}a displays these three dynamical regimes in the $\alpha-\beta$ parameter space. For small values of the damping term $\alpha$, ensured by keeping finite inertia, the homoclinic bifurcation curve is seen to be approximated by a straight line Fig.~\ref{fig : figure2}a. Upon implementing Melnikov's method, \cite{strogatz2018nonlinear,guckenheimer2013nonlinear} the equation of the straight line comes out to be $\beta = \frac{4}{\pi} \alpha$. In conclusion, we see the presence of three different dynamical regimes, namely a limit-cycle regime ($\beta > 1$), a bi-stable regime which can be approximated by ($\frac{4}{\pi} \alpha < \beta \leq 1$), and a fixed point regime approximated by ($\beta \leq \frac{4}{\pi} \alpha$) \cite{strogatz2018nonlinear}.

The bi-stable region is responsible for hysteresis in systems governed by equations like Eq.~\ref{dimensionless mean field equation}. Hence, following \cite{tanaka1997first}, instead of studying the system in its full generality, we break down the self-consistency analysis for our model into forward $\left(f\right)$ and backward $\left(b\right)$ processes. In the forward process, we start from a small $K_1$ value, and therefore, the system is in an incoherent state $\left(r_{1}\approx 0\right)$. This leads to high $\alpha$ and $\beta$ values, indicating that the oscillators are in the limit cycle regime. As we adiabatically increase $K_1$, the oscillators stay in the basin of attraction of the stable limit cycle even after crossing $\beta = 1 (\omega = q)$ and fall into the locked cluster only after $\beta \approx \frac{4}{\pi}$ $\alpha (\omega \approx \frac{4}{\pi} \sqrt{\frac{q}{m}})$, below which the limit cycle vanishes. For the backward process, we start from a high $K_1$ value, and hence the oscillators exist in the fixed point regime, i.e., the oscillators are locked in a cluster and therefore, the system is in a coherent state $\left(0<< r_1 < 1\right)$. As we adiabatically decrease $K_1$, the oscillators remain in the basin of attraction of the stable node until  $\beta = 1$, when the stable node vanishes via a saddle-node bifurcation. Thus, in the backward process, oscillators having $|\omega| \le q = \omega_{b}$ contribute to the locked oscillators, while in the forward process, only those with $|\omega| \leq \frac{4}{\pi} \sqrt{\frac{q}{m}} = \omega_{f}$ are in a locked state and all the oscillators with $\omega > \omega_{f,b}$ drift around the locked cluster. We point out that $K_{2}$ is concealed in $q$ and directly affects the fraction of oscillators in a locked or drifting state. 

The contribution of the locked oscillator($r_{p}^{l}$) to overall coherence for the forward/backward process can now be calculated as $r_{p}^{l} = \int_{-\omega_{f,b}}^{\omega_{f,b}}e^{ip\sin^{-1}(\frac{\omega}{q})}g(\omega)d\omega $. The imaginary part of $r_{p}^{l}$ is zero as $g(-\omega) = g(\omega)$. Hence taking only the real part and noting that  $\theta_{f,b} = \sin^{-1}(\omega_{f,b}/q)$, we arrive at the expression for $r_{p}^{l}$ as follows,
\begin{equation}\label{locked term}
    \begin{split}
    r_{p}^{l} = q\int_{-\theta_{f,b}}^{\theta_{f,b}}\cos(\theta)\cos(p\theta)g(q\sin(\theta))d\theta.
    \end{split}
\end{equation}
The contribution to overall coherence from the drifting oscillators can be accounted for by calculating  $r_{p}^{d} = \int_{|\omega|>\omega_{f,b}}\int_{-\pi}^{\pi} e^{ip\theta}\rho_{d}(\theta,\omega)g(\omega)d\omega d\theta$ where $\rho_{d}(\theta,\omega)$ is the density of drifting oscillator which satisfies $\rho_{d}(\theta,\omega) \propto 1/|\dot\theta|$ \cite{tanaka1997first}. The normalization condition for $\rho_{d}(\theta,\omega)$ gives, $ \int_{-\pi}^{\pi}\rho_{d}(\theta,\omega) d\theta= \int_{0}^{T}\rho_{d}(\theta,\omega)\dot{\theta}dt = 1$ (for a given $\omega$), where $T$ is the time period of the limit cycle solution. Hence we end up with the relation $\rho_{d}(\theta,\omega) = \frac{1}{|\dot{\theta}|T}$, which when plugged into the form of $r_{p}^{d}$ gives us,
\begin{equation}\label{drift1}
    \begin{split}
     r_{p}^{d}  = \int_{|\omega|> \omega_{f,b}} \left[\frac{1}{T}\int_{0}^{T}e^{ip\theta}dt\right]g(\omega)d\omega .
     \end{split}
\end{equation}
To calculate $r_p^d$, we first need to obtain an approximate analytic expression for the limit cycle solution of Eq.~\ref{dimensionless mean field equation}. We follow the method specified in \cite{gao2018self} of writing $\dot{\theta}$ as a Fourier series in  $\theta$ by only considering the first harmonics $(\dot{\theta}(\theta) = A_{0} + A_{1}\cos(\theta)+B_{1}\sin(\theta))$. On substituting this in  Eq.~\ref{dimensionless mean field equation}, we find the expression of the coefficients in terms of $\alpha$($=\frac{1}{\sqrt{qm}}$) and $\beta$($=\frac{\omega}{q}$) such that the first harmonic vanishes, giving us,

\begin{equation}\label{thetadotlc}
\dot{\theta}(\theta) = \frac{\beta}{\alpha} + \frac{\alpha^{2}}{\alpha^{4} + \beta^{2}}\left[\frac{\beta}{\alpha}\cos(\theta) - \alpha \sin(\theta)\right],
\end{equation}

and $\theta(t,\omega)$ by integrating Eq.~\ref{thetadotlc} with time \cite{gao2018self}. As it turns out that $\theta(t,-\omega) = -\theta(t,\omega) $, and $g(-\omega) = g(\omega)$, the imaginary part in Eq.~\ref{drift1} is zero. Thus,
\begin{equation}\label{drift term}
     r_{p}^{d} = \int_{|\omega|> \omega_{f,b}}\left<\cos(p\theta)\right>g(\omega)d\omega .
\end{equation}
 The expression for $\left<\cos(p\theta)\right>$ (for $p \in \{1,2\}$) can now be readily calculated as $\left<\cos(p\theta)\right> = \frac{1}{T}\int_{0}^{T}\cos(p\theta)dt = \int_{0}^{2\pi}\frac{\cos(p\theta)}{\dot{\theta}}d\theta \text{\LARGE $/$} \int_{0}^{2\pi}\frac{1}{\dot{\theta}}d\theta$ to obtain,
     
\begin{equation} \label{cos}
    \begin{aligned}
        \left<\cos(\theta)\right> &= \frac{\beta}{\alpha} \left[ \sqrt{\frac{\beta^2}{\alpha^2} - \frac{\alpha^2}{\beta^2 + \alpha^4}} - \frac{\beta}{\alpha} \right],\\
        \left<\cos(2\theta)\right> &= \left[\frac{\beta^2 - \alpha^4}{\beta^2 + \alpha^4}\right] \times\\
        &\left[ \frac{2\beta(\beta^2 + \alpha^4)}{\alpha^3} \left(\frac{\beta}{\alpha} - \sqrt{\frac{\beta^2}{\alpha^2} - \frac{\alpha^2}{\beta^2 + \alpha^4}} \right) - 1\right].
    \end{aligned}
    \nonumber
\end{equation}

We are now finally ready to write down the self-consistent equations that let us describe the steady state of the coupled oscillator system governed by Eq.~\ref{Gen_KM_hoi}. For the remainder of the work, we consider the intrinsic frequency to be derived from Lorentz distribution, $g(\omega) = \frac{1}{\pi}\frac{1}{1 + \omega^{2}}$ centered around zero. Noting that the integrands in Eqs.~\ref{locked term} and \ref{drift term} for $p \in \{1,2\}$ are even functions, we arrive at,

\begin{equation} \label{r1 self consistent final}
\begin{split}
r_{p}=2q\int_{0}^{\theta_{f,b}}\cos(\theta)\cos(p\theta)g(q\sin(\theta))d\theta  \\
+ 2\int_{\omega_{f,b}}^{\infty}\left<\cos(p\theta)\right>g(\omega)d\omega.
\end{split}
\end{equation}

These two equations together describe the steady-state behavior of the system. 
\section{Numerical Results}
\label{section 4}
We numerically solve the set of self-consistent equations, Eq.~\ref{r1 self consistent final} to find the nontrivial branch of solutions for the order parameter (both forward and backward processes). Fig.~\ref{fig: figure1}a provides a schematic representation of the synchronization profiles of our result in comparison to previously explored models \cite{kuramoto1975self,tanaka1997first,skardal2020higher}. Fig.~\ref{fig: figure1}b presents analytical and simulation results for the $r_{1}$ vs. $K_{1}$ curves for $(m, K_{2}) = (1,1)$ and $(m, K_{2}) = (3,7)$. As for the simulation protocol, we simulate Eq.~\ref{mean field equation} on a network of $N = 10^{4}$ nodes by splitting it into a pair of first-order differential equations and integrating them using the Runge-Kutta 4 algorithm (time-step 0.1). For a chosen value of $m$ and $K_2$, we start with random initial conditions for $\theta (\in [0,2\pi))$ and $\dot\theta (\in [-1,1])$ and $K_1 = 0$. We adiabatically increase $K_1$ in steps of $\Delta K_1$ ($=0.1$, unless specified otherwise) till $K_{1} = 12$ is reached (forward), followed by an adiabatic decrease till $K_1 = 0$ (backward). By adiabatic increase/decrease, we imply that for every $K_1$ except the first ($K_1 = 0$), the initial conditions are taken as the final state obtained for the previous $K_1$ value. At all coupling strengths $K_1$, the order parameter values are calculated after discarding transients by averaging over the steady state.

Fig.~\ref{fig: figure1}b displays a good agreement between the simulation and analytical results. For the forward process, as $K_{1}$ is increased from zero, the system undergoes a first-order phase transition from incoherent to coherent state at a finite critical coupling value ($K_{1}^{f}$). However, for the backward process, the system undergoes abrupt desynchronization at a value 
 ($K_{1}^{b}$), which is less than $K_{1}^{f}$. Hence, hysteresis is observed where the system stays in two different states depending on the initial configuration. The derived self-consistency equations can also be used with other extended-tailed distributions like the Gaussian distribution. In the backward process, there exists a discrepancy between analytical and numerical values (Fig.~\ref{fig: figure1}b) as also reported in \cite{tanaka1997first}. This happens because the maximum value of the $K_1$ chosen for the simulations is 12, when the system starts from a partially coherent state. Whereas, the analytical solutions are for when the system starts from a fully coherent state. Hence, a better fit between analytical and numerical values for the backward process can be obtained by increasing the maximum value of $K_1$ in the simulation protocol. We point out that when $m$ and $K_{2}$ values are both increased, $K_{1}^{f}$ shifts to the right while $K_{1}^{b}$ shifts to the left, revealing a prolonged hysteresis region as illustrated in Fig.~\ref{fig: figure1}b. 

\begin{figure*}[t]
\begin{subfigure}{0.4\textwidth}
    \includegraphics[width=\linewidth]{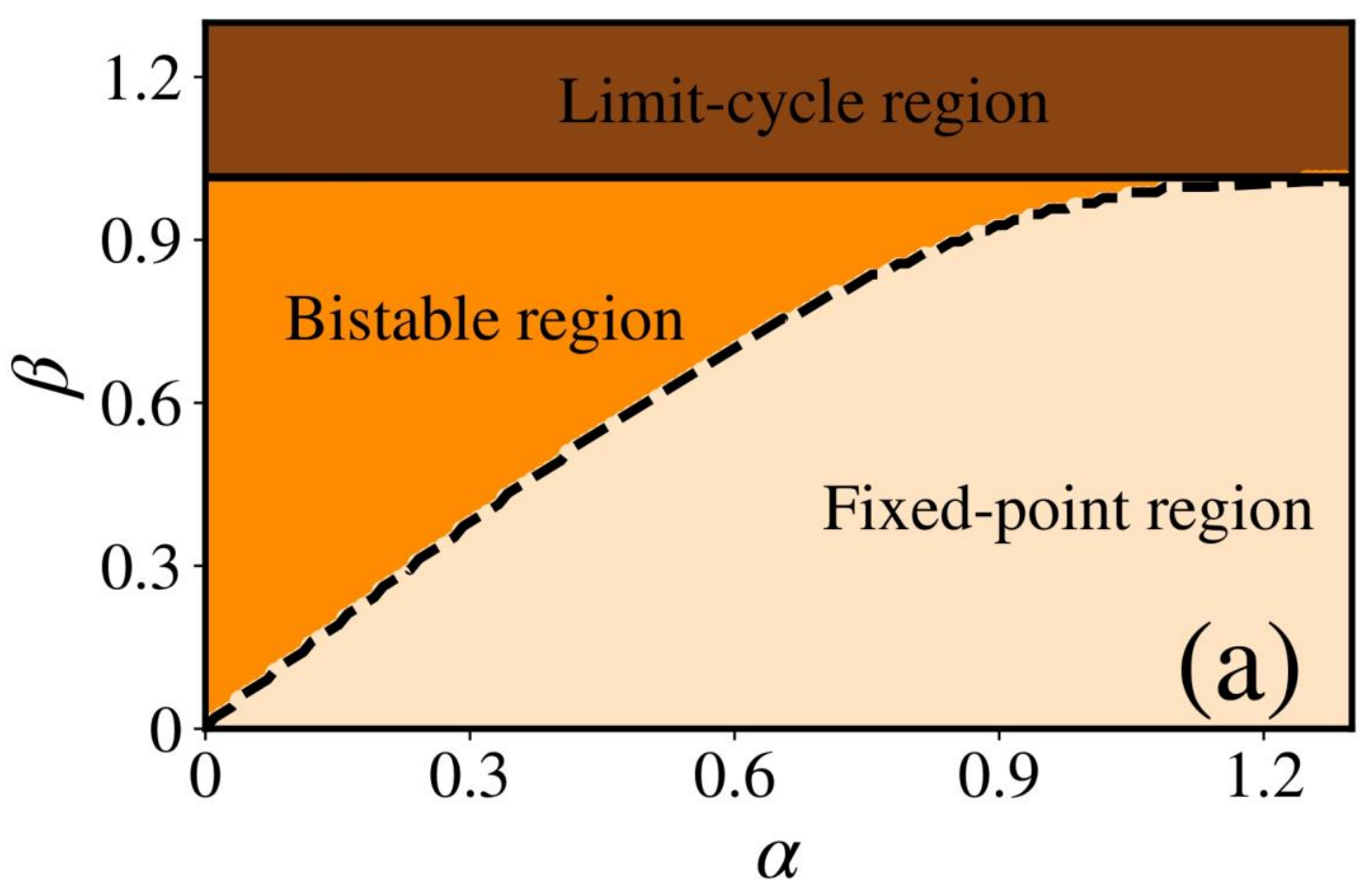}
\end{subfigure}
\begin{subfigure}{0.4\textwidth}
    \includegraphics[width=\linewidth]{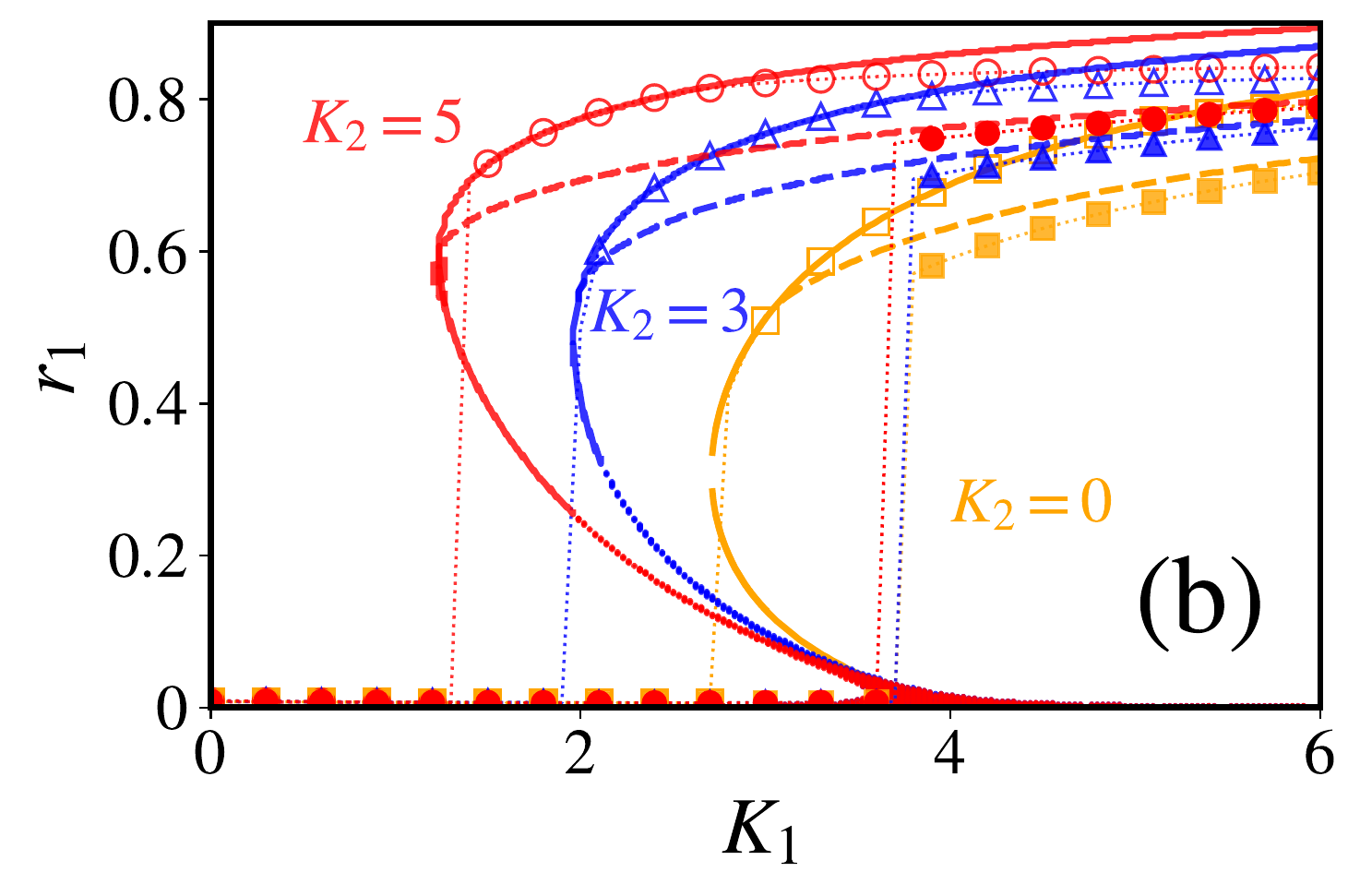}
\end{subfigure}
\begin{subfigure}{0.4\textwidth}
    \includegraphics[width=\linewidth]{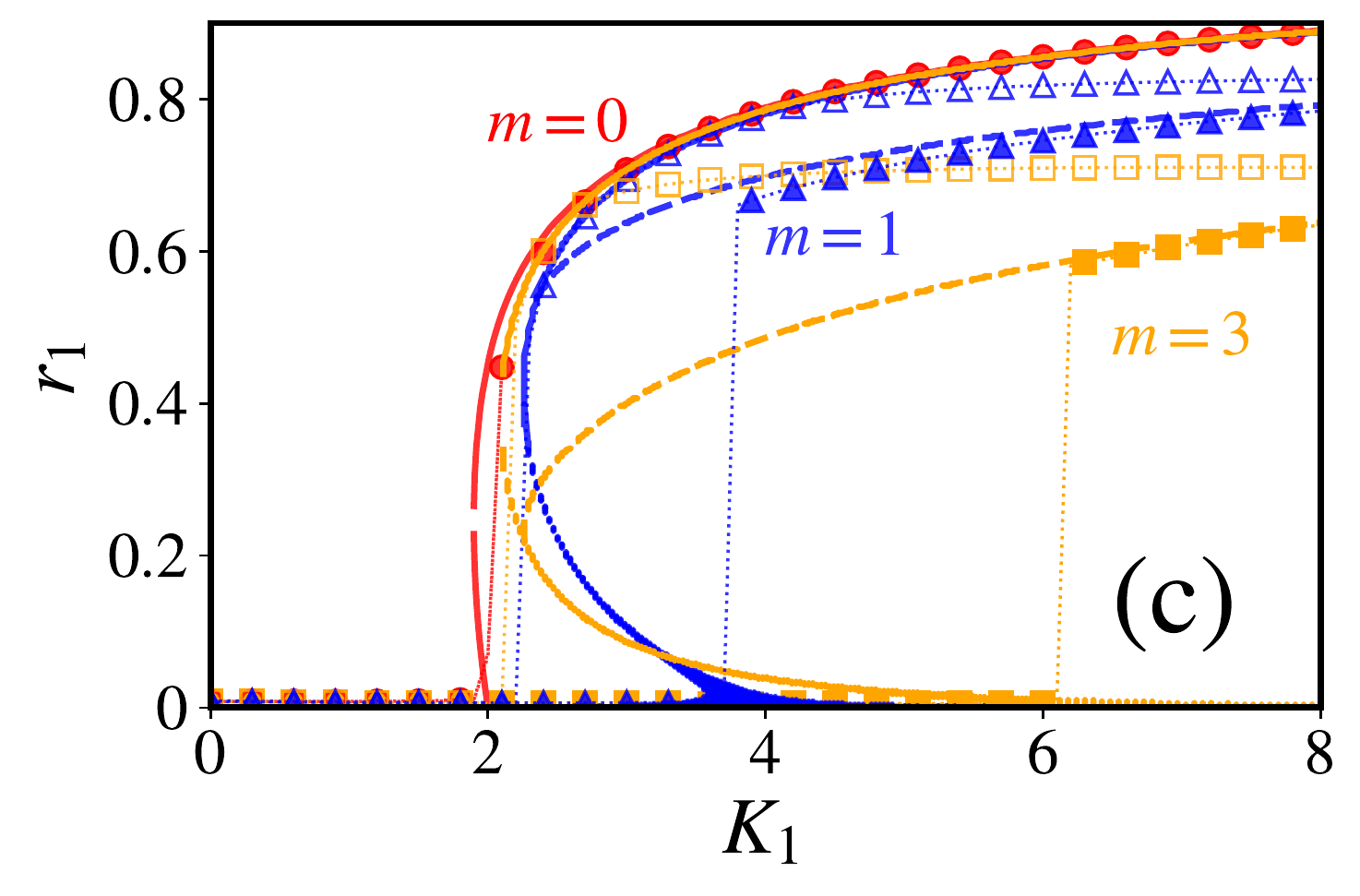}
\end{subfigure}
\begin{subfigure}{0.4\textwidth}
    \includegraphics[width=\linewidth]{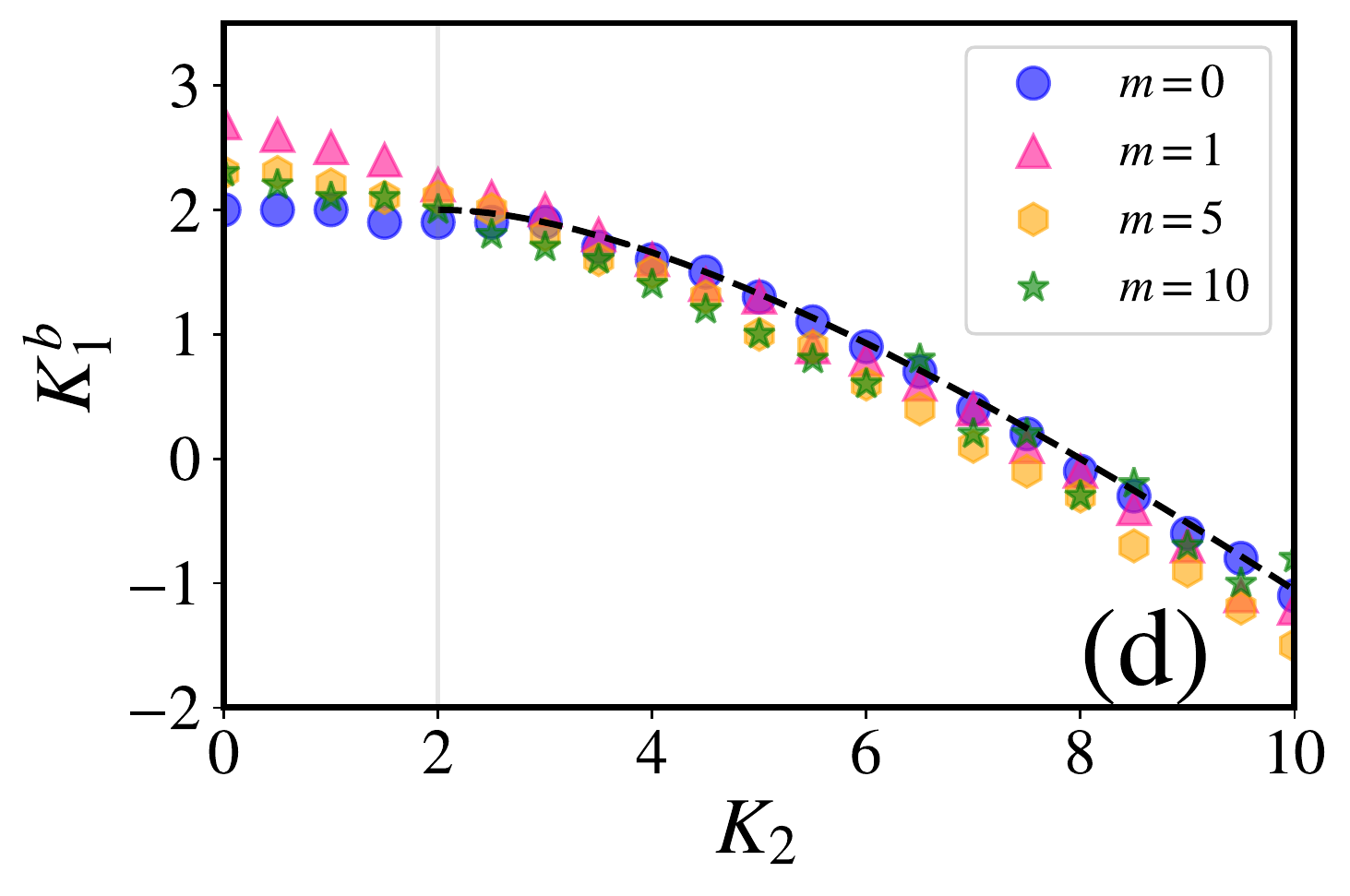}
\end{subfigure}
\caption{(Color online) a) $\alpha(=1/\sqrt{qm})$-$\beta(=\omega/q)$ parameter space. Different dynamical regimes are present in the $\dot{\theta}$ vs. $\theta$ phase space of Eq.~\ref{dimensionless mean field equation}. b)  Synchronization profile $r_1$ versus $K_1$ for $m = 1$ and different values of $K_2 = $ 0 (orange squares), 3 (blue triangles), and 5 (red circles). c) Synchronization profile for a fixed value of $K_2 = 2$ and different values of $m = $ 3 (orange squares), 1 (blue triangles), and 0 (red circles). In both b) and c), the filled and hollow symbols indicate the simulation results for the forward and backward cases, respectively. The dashed and continuous curves represent the analytically calculated values for the forward and backward processes, respectively. d) Backward transition points. The dashed curve represents the analytical predictions of $K_1^b$ for $m=0$ and different $K_2$ ($K_{1}^{b} = 2\sqrt{2K_{2}} - K_{2}$ as derived in Section \ref{Appendix A} for $K_{2}\geq 2$). The scatter plots are the $K_1^b$ vs. $K_2$ for different $m$ = 0, 1, 5, 10 obtained via numerical simulation.}
\label{fig : figure2}
\end{figure*}

 A natural question would then be to address the dependency of the forward and backward transition points on  $m$ and $K_{2}$. To analytically obtain the expression for ($K_{1}^{f}$), we evaluate Eq.~\ref{r1 self consistent final} in the limit $r_{1}\rightarrow 0^{+}$ ($q\rightarrow 0^{+}$). As we take this limit, we see that $\beta/\alpha$($=\omega\sqrt{\frac{m}{q}}$) tends to  very high value as compared to $\alpha^2/(\beta^2 + \alpha^4)$($= \frac{qm}{1+\omega^{2}m^{2}}$). This allows us to perform a Taylor series expansion of $\left<\cos(\theta)\right>$ for $\epsilon = \alpha^2/(\beta^2 + \alpha^4) << 1$ which gives, $\left<\cos(\theta)\right> = \frac{-\alpha^2}{2(\beta^2 + \alpha^4)} + \mathcal{O}(\epsilon^4) \approx \frac{-mq}{2(1+m^2\omega^2)}$. However, in the limit $r_{1}\rightarrow 0^{+}$, $r_{2}\rightarrow 0^{+}$  and the parameter $\alpha \rightarrow \infty$ implying that the limit of the integrals for the forward and backward processes become the same as there exists no bistability region in the parameter space. Taking $\theta_{f,b} = \frac{\pi}{2}$, dividing both sides of Eq.~\ref{r1 self consistent final} by $q$, and evaluating the limit (at which the two equations in Eq.~\ref{r1 self consistent final} decouple) we have, 

 \begin{equation}
 \frac{1}{K_{1}^{f}}= \frac{\pi}{2}g(0) - m\int_{0}^{\infty}\frac{1}{1+m^{2}\omega^{2}}g(\omega)d\omega.
 \nonumber
 \end{equation}
 
 After evaluating the integral and rearranging the terms, we end up with $K_{1}^{f} = 2(m+1)$. We see that the forward transition point is independent of $K_2$ and purely depends on $m$ and, hence, matches the previously derived value of the forward transition point in \cite{gupta2014non,rodrigues2016the}. Fig.~\ref{fig : figure2}b illustrates the effect of varying $K_2 (0.0,3.0,5.0)$ for the case of fixed $m (= 1)$. As expected, $K_1^f$ remains the same for all three cases, validating our analytical result. At this $K_1^f (= 4)$, the magnitude of the first-order jump for fixed $m$ increases with the value of $K_2$. In Fig.~\ref{fig : figure2}c, we study the effect of varying mass ($0.0, 1.0, 3.0$) for the case of fixed $K_2( = 2.0)$. As inertia increases, $K_1^f$ shifts to higher values as predicted analytically. However, we note that the analytically calculated values of $K_1^f$ do not match exactly with numerical simulations owing to the finite size effects. A detailed study has been done in \cite{sb2014}. 

A fairly good analytical approximation for $K_{1}^{b}$, as also pointed out in \cite{sb2014}, would be to obtain the minimum value of $K_{1}$ along the non-trivial branch of the backward self-consistent curve. The simulation results in  Fig.~\ref{fig: figure1}b and Fig.~\ref{fig : figure2}b and \ref{fig : figure2}c are seen to back up this observation for our model. However, obtaining a clean analytical expression for the same by calculating $\frac{dK_{1}}{dr_{1}} = 0$ is not possible because of the complexity of the integrand of the drift oscillator contribution in $r_2$. Alternatively, we resort to simulation results to decipher the dependency of $K_{1}^{b}$ on $m$ and $K_{2}$. From Fig.~\ref{fig : figure2}b, it can be seen that for the backward process, the coherent branch persists till increasingly smaller values of $K_1$ with an increase in the $K_2$ value, after which the system undergoes an abrupt transition to the incoherent state. Hence, it is clear that an increase in $K_{2}$ leads to a decrease in $K_{1}^{b}$. To study the effect of mass on $K_{1}^{b}$, we fix $K_{2}$ and vary $m$ as in Fig.~\ref{fig : figure2}c. It is observed that the backward branches for fixed $K_{2}$(=2) and different $m$ values merge for high $K_1$ values and get separated for low $K_1$ values. As there is an influence of $m$ on the nature of the curve for low $K_1$ values, this indicates the possibility of dependency of $K_{1}^{b}$ on $m$. 

It was shown in \cite{sb2014} that for the pure dyadic case ($K_{2} = 0$), $K_{1}^{b}$ decreases with an increase in $m$ and plateaus out for high $m$ values. In Fig.~\ref{fig : figure2}d, we address how this changes with the introduction of finite $K_{2}$. The $K_1^b$ obtained via simulation (performed for $N = 10^{3}$ number of nodes) for values of $K_2$ ranging from 0 to 10 and different values of $m$(0,1,5,10) are plotted. We see that for small values of $K_{2}$ and finite inertia, an increase in the values in $m$ leads to a decrease in $K_{1}^{b}$. However, we point out that for higher values of $K_{2}$, the effect of $m$ on $K_{1}^{b}$ becomes less pronounced, and desynchronization happens at the same value irrespective of mass. An analytical prediction of $K_{1}^{b}$ becomes possible following this observation by considering the $m=0$ case. We derive self-consistent equations for this case in Section \ref{Appendix A} and obtain $K_{1}^{b}$ as a function of $K_2$ ($K_{1}^{b} = 2\sqrt{2K_{2}} - K_{2}$ for $K_{2}\geq2$) by finding the minimum value of $K_{1}$ in the self-consistency curve. These analytically calculated $K_1^b$ values for the $m=0$ case are represented by the dashed line in Fig.~\ref{fig : figure2}d. It can be clearly observed that for higher values of $K_2(\geq 2)$, the analytical predictions of $K_1^b$ match closely with the ones obtained via simulation for different masses.
\section{Conclusion and Discussion}
\label{section 5}
In this article, we have put forward a generalized analytical framework to study the steady-state behavior of coupled oscillator systems with inertia interacting via higher-order interactions. The analytical predictions, backed up by numerical simulation, show a prolonged hysteretic first-order phase transition to an (in)coherent state. We show that the forward transition point increases linearly with $m$ and is independent of $K_2$. Meanwhile, the backward transition point decreases with $K_2$ and is independent of $m$ for high $K_2$ values. 

 We also highlight here the analytical challenges emanating from combining both the inertia and higher-order interaction terms in the Kuramoto model. For the coupled Kuramoto oscillator model, two widespread analytical techniques are the self-consistency method and the Ott-Antonsen (OA) approach to studying the synchronization profile of the entire system. OA dimensionality reduction method provides a very easy way to obtain the time evolution form of $r_1$, solving which yields a complete description of bifurcations leading to synchronization. While the OA approach was successfully extended for the Kuramoto model with higher-order interactions, for the second-order Kuramoto models, the density function containing phase term also depends on velocity,  posing restrictions on practicing the OA method. Hence, the Kuramoto model with inertia has been analytically solved using the self-consistency method.
The challenge in this method is first finding the limit of integration in the self-consistent equation by finding an approximate frequency bound for the oscillators, which are in the locked state as a function of external parameters and the order parameter. Further, the drifting oscillators also contribute to the overall coherence, which needs to be accounted for, unlike in the first-order Kuramoto model, where it is zero. Here, we have employed the self-consistency method to predict the steady-state behavior of the Kuramoto oscillators having both inertia and higher-order terms.

We have presented the results for triadic interactions; however, extending our analysis to other powers of higher-order interactions is easy, as long as the sinusoidal coupling function contains $\theta_i$ term only. For example, the detailed analysis of quartic interactions is presented in Section \ref{Appendix B}. Further, developing the self-consistent method for other choices of higher-order coupling functions, such as $\sin(\theta_j + \theta_k - 2\theta_i)$ \cite{tanakat2011multistable, skardal2019abrupt} along with pairwise coupling proves to be complicated because of the existence of higher order harmonics in the mean-field equation; however, the self-consistency approach can work for this form of the coupling in the absence of pairwise coupling  (pure triadic case) which we have explored in Section \ref{Appendix c}. 
An immediate future direction of our work would be to extend our analysis to diluted simplicial complexes, which can provide fundamental insights into the dynamics of various real-world complex systems such as power grids.

\begin{acknowledgments}
SJ gratefully acknowledges SERB Power grant SPF/2021/000136, and thanks  Mehrnaz Anvari and Baruch Barzel for useful comments on the manuscript.
\end{acknowledgments}

\section{Appendix}

\subsection{Derivation for $m = 0$}
\label{Appendix A}
\setcounter{figure}{0}
\setcounter{equation}{0}
\renewcommand{\thefigure}{A-\arabic{figure}}
\renewcommand{\theequation}{A-\arabic{equation}}

This section elaborates on the self-consistency analysis and the derivation for the forward and backward transition points for $m = 0$. In this case, a change from a smooth (second-order) transition to synchronization to an abrupt (first-order) one, along with hysteresis, is observed with an increase in the $K_2$ value. The occurrence of hysteresis can be accounted for by the shift of the backward transition point to lower $K_1$ values with an increase in $K_2$. Note that the results for the first-order Kuramoto model with higher-order interactions were reported in \cite{skardal2020higher} using the Ott-Antonsen dimensionality reduction method. Here, we derive the closed-form solution for the bifurcation curve using the self-consistency method.

The dynamical equation for our model takes the following form.
\begin{equation}\label{m0 mean field equation}
        \dot{\theta}_{i} =  \omega_i + \frac{K_1}{N}\sum_{j=1}^{N}\sin(\theta_j-\theta_{i}) + \frac{K_2}{N^2}\sum_{j=1}^{N} \sum_{k=1}^{N} \sin(2\theta_j-\theta_k -\theta_{i})
\end{equation}
Following the same procedure as in the article, the mean-field equation in the rotating frame is obtained as $\dot{\theta_i} = \omega_i - q\sin(\theta_i)$, where $q = r_{1}(K_{1} +K_{2}r_{2})$. All the oscillators with intrinsic frequency $|\omega_i|\leq q$ go to the fixed point state, while others are in a drift state. Unlike in the finite-inertia case, here the limit of the integrals for forward and backward cases become the same ($\theta_{f} = \theta_{b} = \frac{\pi}{2}$), enabling us to study the system in full generality. In the thermodynamic limit, the locked oscillator contribution has the same form as derived in the article,

\begin{equation}\label{appendix locked term}
    \begin{split}
    r_{p}^{l} = q\int_{-\pi/2}^{\pi/2}\cos(\theta)\cos(p\theta)g(q\sin(\theta))d\theta,
    \end{split}
\end{equation}
which upon integrating for $p \in \{1,2\}$ gives us,

\begin{equation}\label{locked m = 0}
    \begin{aligned}
        r_{1}^{l} &= \dfrac{\sqrt{q^{2}+1} - 1}{q},\\
        r_{2}^{l} &=  \frac{2}{\pi q^{2}}\left[(q^{2} + 2)\tan^{-1}(q) -2q\right].
    \end{aligned}
\end{equation}
{The contribution of drifting oscillators to the global order parameter, however, has the following expression,}

\begin{figure}[t]
  \centering
  \includegraphics[width=0.49\textwidth, height=0.32\textwidth]{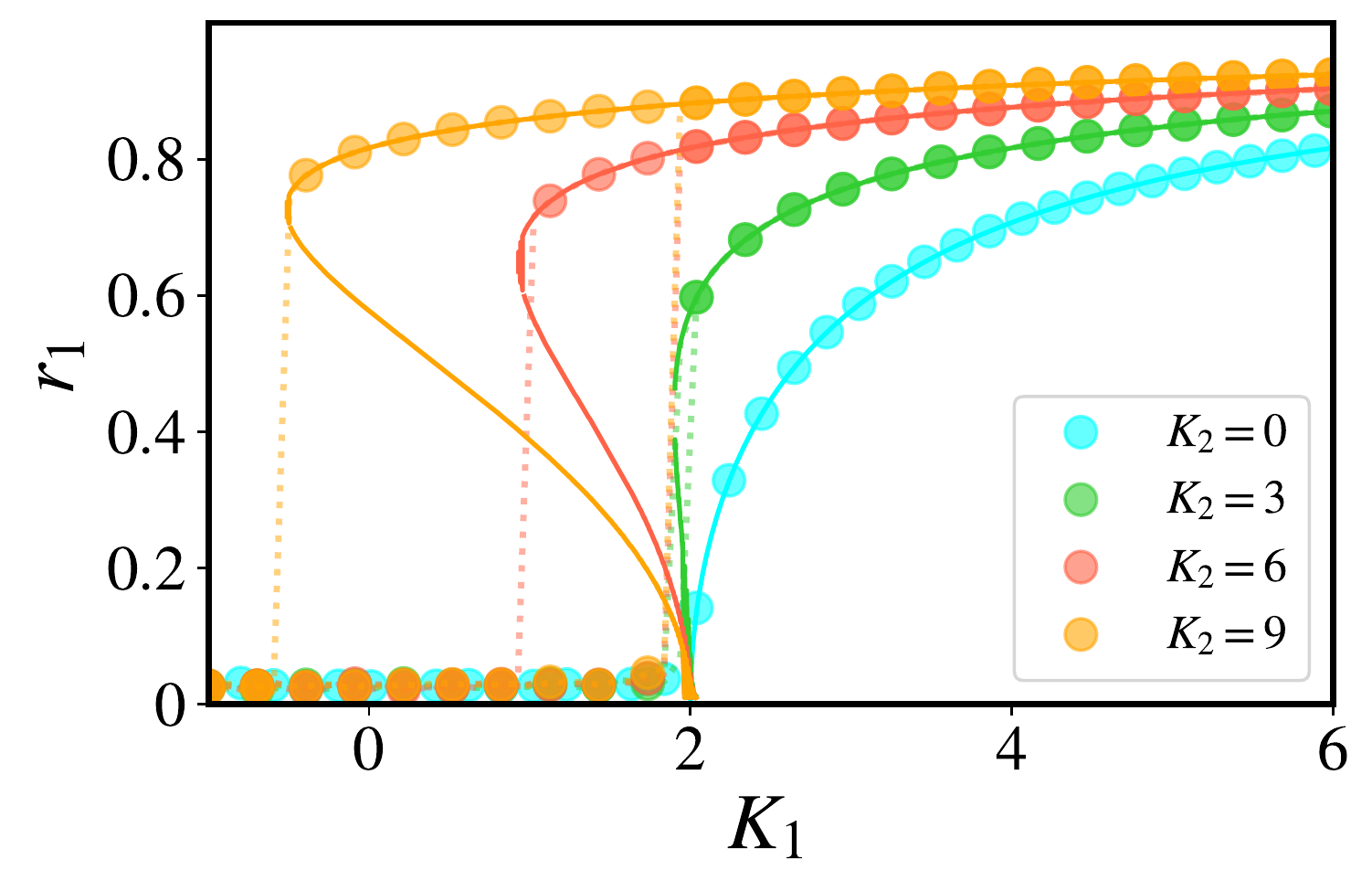}
  \caption{Synchronization profile of Eq.~\ref{m0 mean field equation}, for and $K_2 = 0$ (cyan), $3 $ (green), $6$ (red), and $9$ (yellow). The circles represent simulation results for the forward and backward cases, while the continuous line represents the analytically predicted values.}
  \label{fig: SM-m0}
\end{figure}

\begin{equation*}
    \begin{aligned}
    r_{p}^{d} &= \int_{|\omega|>q}\int_{-\pi}^{\pi} e^{ip\theta}\rho_{d}(\theta,\omega)g(\omega)d\omega d\theta,\\
    r_{p}^{d} &= \int_{|\omega|>q}\int_{-\pi}^{\pi}e^{ip\theta}\frac{\sqrt{\omega^{2} - q^{2}}}{2\pi|\omega - q\sin(\theta)|}g(\omega)d\omega d\theta . \\
    \end{aligned}
\end{equation*}

The expression for the density of the drifting oscillators is derived by noting that $\rho_{d}(\theta,\omega) \propto 1/|\dot{\theta}| = c/|\omega - q\sin(\theta)|$, where $c$ is the normalization constant such that $\int_{-\pi}^{\pi}\rho_{d}(\theta,\omega)d\theta = 1$. The value of $r_{1}^{d}$ is zero since $\rho_{d}(\theta+\pi,-\omega) = \rho_{d}(\theta,\omega)$ and $g(-\omega) = g(\omega)$. Meanwhile, in $r_{2}^{d}$, only the imaginary term is zero, since $\rho_{d}(-\theta,-\omega) = \rho_{d}(\theta,\omega)$ and $g(-\omega) = g(\omega)$ while the real part is not zero. Since the integrand is even, we can evaluate the integral as follows:

\begin{equation}\label{drifting m=0}
    \begin{aligned}
       r_{2}^{d} &= 2\int_{q}^{\infty}\left[\int_{-\pi}^{\pi}\cos(2\theta)\frac{\sqrt{\omega^{2} - q^{2}}}{2\pi(\omega - q\sin(\theta))}d\theta \right]g(\omega)d\omega, \\
       &= 2\int_{q}^{\infty}\left[\dfrac{2\omega}{q^2}\left(\sqrt{\omega^{2} - q^{2}}-\omega\right) + 1 \right]g(\omega)d\omega,\\
       & = -\dfrac{2}{\pi q^{2}}\left((q^{2} +2)\tan^{-1}(q) - 2q\right) + \dfrac{2+q^{2}-2\sqrt{q^{2}+1}}{q^{2}}.
    \end{aligned}
\end{equation}

Using the self-consistency condition $(r_{p} = r_{p}^{l} + r_{p}^{d})$ and Eq.~\ref{locked m = 0}, Eq.~\ref{drifting m=0} we arrive at a set of self-consistent equations which when solved give us the analytical predictions of the steady state behavior,

\begin{subequations}\label{eq:bifurcation equations}
    \begin{equation}\label{eq:m=0 bifurcation}
    \begin{aligned}
        r_{1} &= \dfrac{\sqrt{q^{2}+1} - 1}{q},
    \end{aligned}
    \end{equation}
    \begin{equation}
    \begin{aligned}
        r_{2} &= \dfrac{2+q^{2}-2\sqrt{q^{2}+1}}{q^{2}},
    \end{aligned}
    \end{equation}
\end{subequations}

where $q = r_{1}(K_{1} + r_{2}K_{2})$. Solving Eq.~\ref{eq:bifurcation equations} gives us the relation $r_{2} = r_{1}^{2}$, which, when plugged back into Eq.~\ref{eq:m=0 bifurcation} leads to the closed form solution for the bifurcation curve, 

\begin{equation}\label{final version of the self consistency}
    K_{1} = \dfrac{2}{1-r_{1}^{2}} - r_{1}^{2}K_{2}.
\end{equation}

We point out that this equation matches with the one obtained by \cite{skardal2020higher} using the OA method.

Fig.~\ref{fig: SM-m0} presents analytical and simulation results for multiple values of triadic coupling strength $K_{2}$. As predicted by our analysis, the forward synchronization transition happens at $K_1^f = 2$, while the backward transition point shifts to lower values of $K_1$ as $K_2$ increases. From Fig.~\ref{fig: SM-m0},  we notice that the minima of the self-consistency curve fairly well approximates the backward de-synchronization point. To this end, setting $\frac{dK_{1}}{dr_{1}} = 0$, we obtain $r_{1}^{b} = \sqrt{1 - \sqrt{\frac{2}{K_{2}}}}$ for $K_{2} \geq 2$. Plugging the form of $r_{1}^{b}$ back into Eq.~\ref{final version of the self consistency} gives $K_{1}^{b} = 2\sqrt{2K_{2}} - K_{2}$. This is used in Fig.~\ref{fig : figure2}d in the article.

\subsection{Extension to quartic interactions}
\label{Appendix B}
\setcounter{figure}{0}
\setcounter{equation}{0}
\renewcommand{\thefigure}{B-\arabic{figure}}
\renewcommand{\theequation}{B-\arabic{equation}}

\begin{figure*}[t]
\centering
\includegraphics[scale=1]{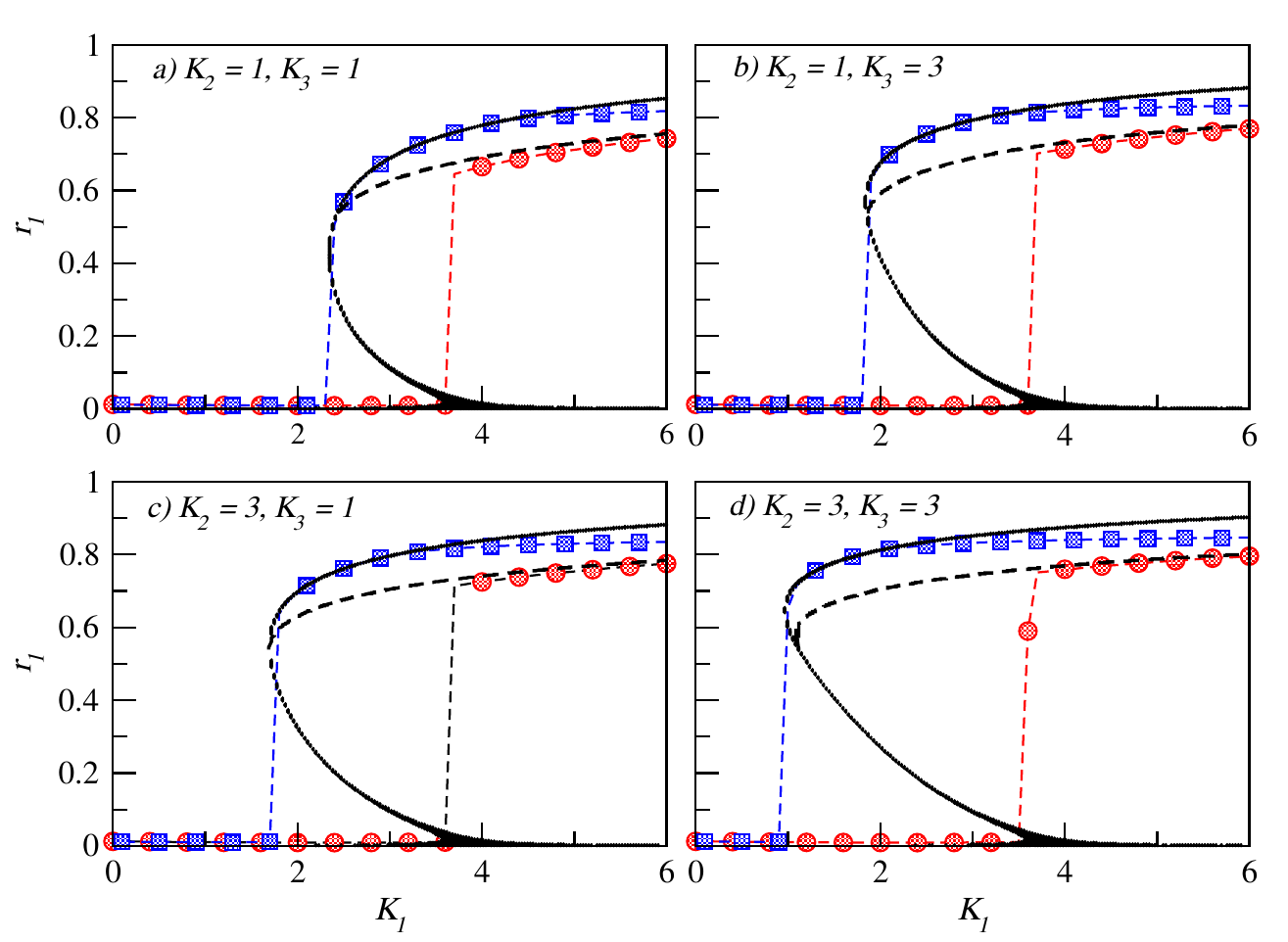}
\caption{Synchronization profiles for the Kuramoto model with inertia involving quartic interactions with $m=1$ and different combinations of $K_2$ and $K_3$ values. The red circles and blue squares represent the simulation results for the forward and backward cases of Eq.~\ref{Gen_KM_quar} for $N=1000$ nodes, respectively. Whereas the dashed and continuous curves are the analytical predictions from \ref{total self consistent} for the forward and backward cases. }
\label{fig: quartic}
\end{figure*}

This section explains how the self-consistency analysis presented in the article can be easily extended to include quartic interactions. We add a quartic interaction term to our model as proposed in \cite{skardal2020higher}. In our study, we find a good agreement between the analytical predictions of our self-consistency analysis and the simulation results. The detailed analysis of the impact of quartic interactions on the dynamics is out of the scope of this discussion. The dynamical equation for our model is given as follows:
\begin{equation}\label{Gen_KM_quar}
    \begin{aligned}
    m\ddot{\theta_i} =  & -\dot{\theta_i} + \omega_i + \frac{K_{1}}{N}\sum_{j = 1}^{N}\sin(\theta_{j} - \theta_{i}) \\
    & +\frac{K_{2}}{N^{2}} \sum_{j = 1}^{N}\sum_{k = 1}^{N}\sin(2\theta_{j} - \theta_{k}- \theta_{i}) \\
    & + \frac{K_{3}}{N^{3}}\sum_{j = 1}^{N}\sum_{k = 1}^{N}\sum_{l = 1}^{N}\sin(\theta_{j} + \theta_{k}-\theta_{l} - \theta_{i}).
    \end{aligned}
\end{equation}

Upon using the definition of the generalized order parameter as defined in the article, we can convert the above equation into a mean-field form as,
\begin{equation*}
\begin{aligned}
     m\ddot{\theta}_{i} &=  -\dot{\theta}_{i} + \omega_i + K_{1}r_{1}\sin(\psi_{1} - \theta_{i})\\ &+K_{2}r_{1}r_{2}\sin(\psi_{2} - \psi_{1}- \theta_{i}) + K_{3}r_{1}^{3}\sin(\psi_{1} - \theta_{i}).
\end{aligned}
\end{equation*}
By moving to the rotating frame, we set $\psi_1$ and $\psi_2$ as zero, which gives $m\ddot{\theta}_{i} =  -\dot{\theta}_{i} + \omega_i - q\sin(\theta_{i})$, where $q = r_{1}(K_{1} +K_{2}r_{2} + K_{3}r_{1}^{2})$ is the overall coupling constant. We proceed exactly, as in the article, to arrive at the self-consistency equations defining the steady-state behavior of the model.

\begin{widetext}
\begin{subequations}\label{total self consistent}
     \begin{equation} \label{r1 self consistent quartic final}
     \begin{split}
     r_{1} = 2q\int_{0}^{\theta_{f,b}}\cos^{2}(\theta)g(q\sin(\theta))d\theta + 2\int_{\omega_{f,b}}^{\infty}    \omega\sqrt{\frac{m}{q}}\left[\sqrt{\omega^{2}\frac{m}{q} - \frac{qm}{1+\omega^{2}m^{2}}} - \omega\sqrt{ \frac{m}{q}}\right]g(\omega)d\omega,
     \end{split}
     \end{equation}
     \begin{equation}\label{r2 self consistent quartic final}
         \begin{split}
              r_{2} &= 2q\int_{0}^{\theta_{f,b}}\cos(\theta)\cos(2\theta)g(q\sin(\theta))d\theta + \\ 
              & 2\int_{\omega_{f,b}}^{\infty}    \left[\frac{\omega^{2} m^{2}-1}{\omega^{2} m^{2}+1}\right]\left[\frac{2\omega}{q}\left(\frac{\omega^{2}m^{2}+1}{\sqrt{qm}}\right)\left(\sqrt{\omega^{2}\frac{m}{q} - \frac{qm}{1+\omega^{2}m^{2}}} - \omega\sqrt{ \frac{m}{q}}\right) - 1\right]g(\omega)d\omega.
         \end{split}
     \end{equation}
\end{subequations}
\end{widetext}
Fig.~\ref{fig: quartic} plots $r_1$ as a function of $K_1$ for different combinations of $K_2$ and $K_3$, and we see that the analytical predictions match the numerical simulation results.
\subsection{Derivation for the $\sin(\theta_j + \theta_k - 2\theta_i)$ model}
\label{Appendix c}
\setcounter{figure}{0}
\setcounter{equation}{0}
\renewcommand{\thefigure}{C-\arabic{figure}}
\renewcommand{\theequation}{C-\arabic{equation}}

In this section, we present the self-consistency analysis for the Kuramoto model with inertia involving purely triadic interactions via the $\sin(\theta_j + \theta_k - 2\theta_i)$ coupling function. Research on the dynamics of Kuramoto oscillators coupled via the said sinusoidal function ($\dot\theta_i = \omega_i + \frac{K_2}{N^2} \sum_{j=1}^{N} \sum_{k=1}^{N} \sin(\theta_j + \theta_k - 2\theta_i )$) \cite{skardal2019abrupt} has shown the presence of cluster formation and a continuum of abrupt de-synchronization transitions based on initial conditions.
Crucially, no synchronization transition has been reported for this model. Surprisingly, our studies report that inertia has no effect on the synchronization profile of this system. We infer that considering purely triadic interactions removes the distinction between inertia-less and finite-inertia cases. 
The dynamical equations for this model are given as:
\begin{equation}\label{MM-HO-M1OG}
    m\ddot{\theta_i} = \omega_i - \dot{\theta_i} + \frac{K_2}{N^2} \sum_{j=1}^{N} \sum_{k=1}^{N} \sin(\theta_j + \theta_k - 2\theta_i).
\end{equation}

Using the definition of the generalized order parameter, we write Eq.~\ref{MM-HO-M1OG} in the mean-field format as,

\begin{equation}\label{MM-Ho-M1MF}
m\ddot{\theta_i} = \omega_i - \dot{\theta_i} + K_2r_1^2\sin(2\psi_1 - 2\theta_i).
\end{equation}
Upon simulating the dynamics of this equation, we observe that analogous to the studies reported in \cite{skardal2019abrupt}, there exists no forward synchronization in the system rather, a sequence of de-synchronization transitions is observed based on the asymmetry in the initial conditions. Thus, we focus the analytical derivation only on the de-synchronization profiles.

\begin{figure}[t]
\centering
\includegraphics[scale=0.7]{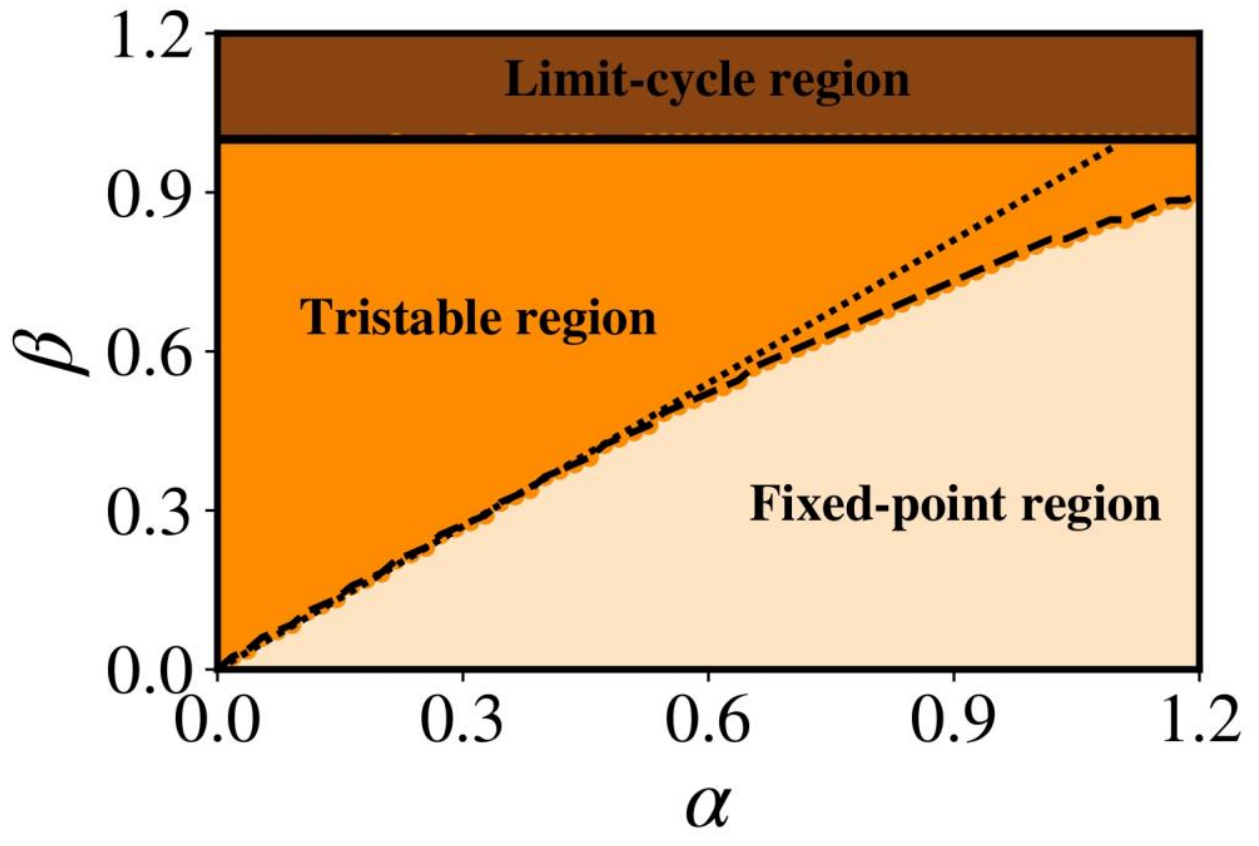}
\caption{The parameter space for Eq.~\ref{MM-HO-M1PS} representing the limit-cycle, tristable, and fixed-point regimes, respectively. The line at $\beta = 1$ represents the saddle-node bifurcation in the system. On the boundary of the tri-stable and fixed-point regimes, the scatter plot indicates the actual homoclinic bifurcation curve while the dotted line is the approximation obtained using Melnikov's method ($\beta = \frac{2\sqrt{2}}{\pi}\alpha$).}
\label{fig: SM-paramspace}
\end{figure}

Dropping the subscript $i$, by going to a suitable rotating frame, we set $\psi = 0$, which gives us $m\ddot{\theta} = \omega - \dot{\theta} - K_2r_1^2\sin(2\theta)$, where the probability distribution g($\omega$) is unimodal and symmetric about the mean zero. To study the steady-state behavior, we invoke a time transformation as $\tau = \sqrt{\frac{K_2r_1^2}{m}} t$, which gives:
\begin{equation}\label{MM-HO-M1PS}
    \ddot\theta = \beta - \alpha\dot\theta - \sin(2\theta),
\end{equation}
where $\beta = \frac{\omega}{K_2r_1^2}$ and $\alpha = \frac{1}{\sqrt{K_2r_1^2m}}$. The parameter space of Eq.~\ref{MM-HO-M1PS} is qualitatively similar to that of the model considered in the article. The quantitative differences between the two are as follows: (i) For each $\beta$ such that $|\beta| \le 1$, we now have two stable fixed points given by $\theta_1^* = \frac{1}{2}\sin^{-1}{\beta}, \,\,\, \text{and} \,\,\, \theta_2^* = \frac{1}{2}\sin^{-1}{\beta} + \pi$ separated by two saddles. (ii) The separatrix equation is given by $\theta(t) =  \sin^{-1}{\tanh(\sqrt{2}t)}, \,\,\, \dot{\theta}(t) = \sqrt{2} \frac{1}{\cosh(\sqrt{2}t)}$. Using this equation of separatrix and implementing Melnikov's method \cite{guckenheimer2013nonlinear}, the equation for homoclinic bifurcation can be obtained as $\beta = \frac{2\sqrt{2}}{\pi}\alpha$. Thus, as seen in Fig.~\ref{fig: SM-paramspace}

 \begin{enumerate}
    \item if $|\beta| > 1$, the system goes to a limit cycle; 
    \item if $|\beta| \le \frac{2\sqrt{2}}{\pi}\alpha$, the system goes to a stable fixed point state, where the choice of the fixed point depends upon the initial conditions. \item Finally, if $\frac{2\sqrt{2}}{\pi}\alpha<|\beta|<1$, we now have a tri-stable state with the simultaneous presence of one stable limit cycle and two stable fixed points. 
    \end{enumerate}

\begin{figure*}[t]
\centering
\includegraphics[scale = 1]{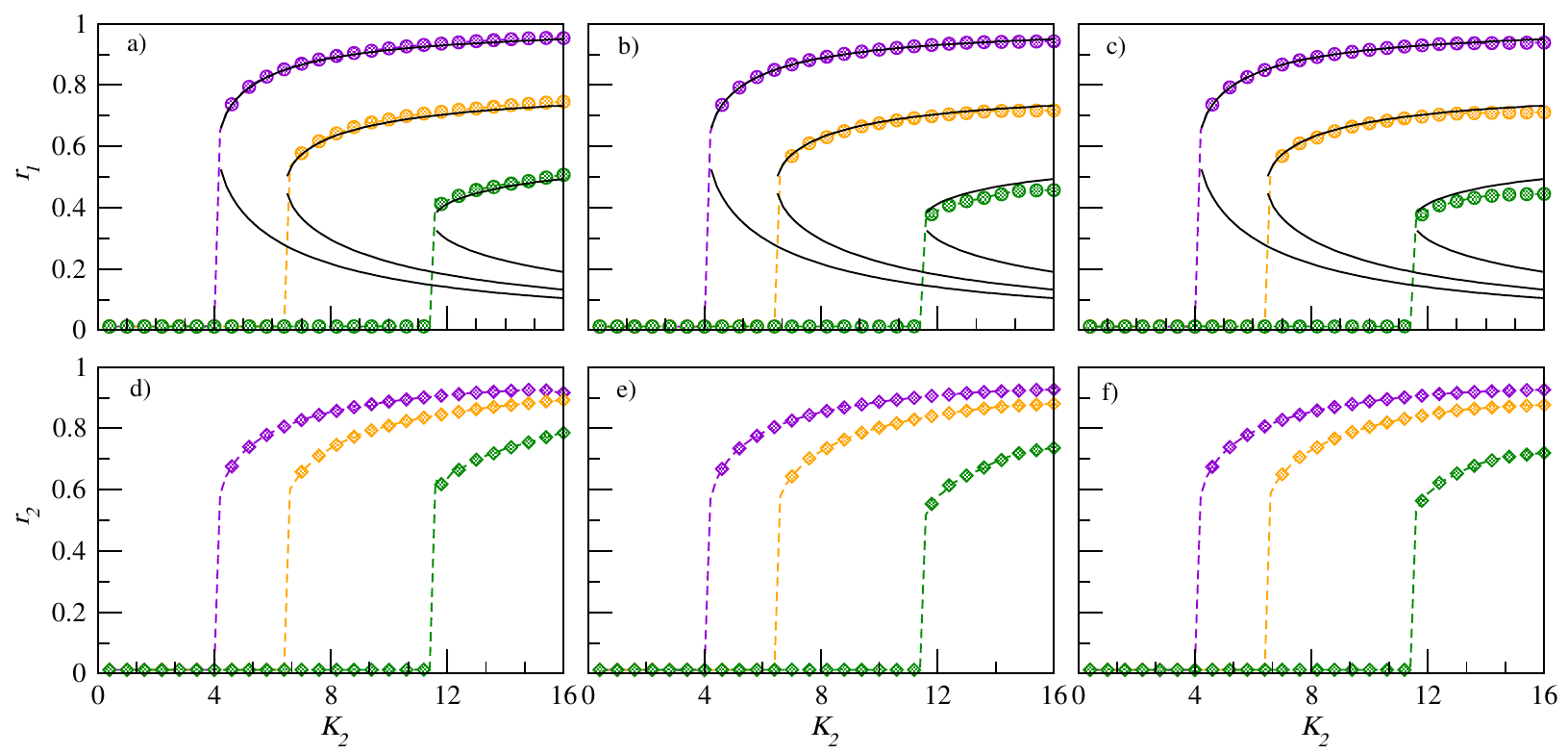}
\caption{ The top panel displays the $r_1$ versus $K_2$ behaviour for Eq.~\ref{MM-Ho-M1MF} for a) $m = 0$, b) $m = 1$, and c) $m = 3$  respectively. The violet, yellow, and green curves represent the de-synchronization profiles for $\eta$ equal to 1, 0.9, and 0.8, respectively. For each case, the continuous curve represents the analytically obtained values. The bottom panel displays the simulation-obtained values for the Daido order parameter ($r_2$) for the corresponding cases (d) $m=0$, e) $m=1$, and f) $m=3$.}
\label{fig: Model1}
\end{figure*}

In terms of $\omega$, $K_2$, and $r_1$, it means that for a particular $K_2$ and corresponding steady-state value of $r_1$, the oscillators with $|\omega| > K_2r_1^2$ become drifting oscillators. The oscillators with $|\omega| \le \frac{2\sqrt{2}}{\pi} \sqrt{\frac{K_2r_1^2}{m}}$ contribute to the formation of two diametrically opposite clusters of locked oscillators. Finally the oscillators with $\frac{2\sqrt{2}}{\pi} \sqrt{\frac{K_2r_1^2}{m}} < |\omega| \le 1$ are in the tri-stable stable region. However, as we will deal only with the system de-synchronization profile, the oscillators in the tri-stable region will also go to their respective fixed points and contribute to cluster formation.

The next step is to calculate the locked and drifting oscillator contribution to the order parameter. First, let us calculate the contribution from the locked oscillators. To account for the two clusters of locked oscillators, we introduce a variable $\eta(\omega) \in [0,1]$. The values of $\eta(\omega)$ and $1-\eta(\omega)$ are the probabilities that an oscillator with intrinsic frequency $\omega$ will go to the fixed points $\theta_1^* = \frac{1}{2}\sin^{-1}{\beta}$ and $\theta_2^* = \frac{1}{2}\sin^{-1}{\beta} + \pi$, respectively. For simplicity, we consider only symmetric cases of the function $\eta(\omega)$, such that $\eta(\omega) = \eta(-\omega)$. The contribution of locked oscillators is then given by $r_1^l = \int_{-K_2r_1^2}^{K_2r_1^2} [(1-\eta(\omega)) e^{i\theta(\omega) + \pi} + \eta(\omega)e^{i\theta(\omega)}] g(\omega)d\omega$. As $e^{i\theta(\omega) + \pi} = - e^{i\theta(\omega)}$ and $\sin(-x) = -\sin(x)$, we have:
\begin{equation}\label{MM-HO-M1rl}
    r_1^l = \int_{-K_2r_1^2}^{K_2r_1^2} (2\eta(\omega) - 1) \cos \theta(\omega) g(\omega) d\omega.
\end{equation}

Next, let us consider the contribution of the drifting oscillators to the order parameter. Let $\rho_d(\theta,\omega)$ be the density of drifting oscillators which satisfies $\int_{-\pi}^{\pi} \rho_d(\theta,\omega) d\theta = 1$. The continuity equation for the conservation of the number of oscillators gives $\rho_d(\theta,\omega) = c/\dot\theta(\theta,\omega)$. An expression for $\dot\theta$ can be obtained by following an analogous method as in the article by considering $\dot\theta = A_0 + A_1 \cos(2\theta) + B_1 \sin(2\theta)$. By substituting this in Eq.~\ref{MM-HO-M1PS}, the expression for $\dot\theta$ can be obtained as:
\begin{equation}\label{MM-Ho-M1v}
\dot\theta(\omega,\theta) = \frac{\beta}{\alpha} + \frac{2 \beta \alpha}{\alpha^4+4\beta^2} \cos(2\theta) - \frac{\beta^3}{\alpha^4 + 4\beta^2}\sin(2\theta).
\end{equation}
Eq.~\ref{MM-Ho-M1v} can be simplified by considering $h_0 = \beta/ \alpha$ and $\frac{e^{ih_2}}{h_1} = 2h_0 + i\alpha$, which gives $\dot\theta = h_0 + h_1\cos(2\theta + h_2)$. By integrating over the normalization condition of $\rho_d(\theta,\omega)$, we get $c = \sqrt{h_0^2 - h_1^2}/2\pi$, which gives the following equation for the density of drifting oscillators:

\begin{widetext}
\begin{equation}\label{MM-HO-M1rho}
    \rho_d(\theta,\omega) = \left| \frac{1}{2\pi} \sqrt{\frac{\omega^2m}{K_2r_1^2} - \frac{K_2r_1^2m}{1+4m^2\omega^2}} \times \frac{1}{\frac{\omega\sqrt{m}}{\sqrt{K_2r_1^2}} + \sqrt{\frac{K_2r_1^2m}{1+4m^2\omega^2}}\cos(2\theta(\omega) + \tan^{-1}(\frac{1}{2m\omega}))} \right | ,
\end{equation}
\begin{equation}\label{r1d}
r_1^d = \int_{|\omega|>K_2r_1^2}\int_{-\pi}^{0} [ e^{i\theta(\omega)} \rho_d(\theta,\omega) + e^{i(\theta(\omega) + \pi)} \rho_d(\theta+\pi,\omega)] g(\omega) d\theta d\omega.
\end{equation}
\end{widetext}

From Eq.~\ref{MM-HO-M1rho}, we notice that $\rho_d(\theta+\pi,\omega) = \rho_d(\theta,\omega)$. Thus, the contribution of drifting oscillators to order parameter $r_1^d = \int_{|\omega|>K_2r_1^2} \int_{-\pi}^{\pi} e^{i\theta(\omega)}\rho_d(\theta,\omega) g(\omega) d\theta d\omega$ can be simplified as Eq.~\ref{r1d}

As $e^{i(\theta(\omega) + \pi)} = -e^{i\theta(\omega)}$, we get $r_1^d = 0$. Therefore, from Eq.~\ref{MM-HO-M1rl} and the self consistency condition ($r_{1} = r_{1}^{l} + r_{1}^{d}$), we get
\begin{equation}\label{MM-m1-rf}
    r_1 = \int_{-K_2r_1^2}^{K_2r_1^2} (2\eta(\omega) - 1) \cos \theta(\omega) g(\omega) d\omega.
\end{equation}

We now compare the results obtained by simulating Eq.~\ref{MM-Ho-M1MF} with the analytical values predicted by the Eq.~\ref{MM-m1-rf} in Fig.~\ref{fig: Model1}. For simplicity, we consider $\eta(\omega)$ to be a constant function with respect to $\omega$. While simulating for a given value of $\eta$, we consider the initial phase of the oscillators as $0$ with probability $\eta$ and $\pi$ with probability $(1-\eta)$. With these initial conditions, we start from $K_2=K_{max} = 16$, and then adiabatically decrease $K_2$ till $K_{min}=0$. At each $K_2$, we integrate Eq.~\ref{MM-Ho-M1MF} using the RK4 algorithm and calculate the value of $r_1$ and $r_2$ by averaging over all the iterations after removing the transients.

We note that due to the vanishing of the drift term, the form of the bifurcation curve Eq.~\ref{MM-m1-rf} is the same as well for the $m=0$ case. Fig.~\ref{fig: Model1} also shows that the $r_1$ and $r_2$ graphs are almost identical for the finite mass and the mass-less case, with identical backward transition points. Hence, considering purely triadic interactions eliminates the distinction between the finite inertia and inertia-less case. For $\eta = 0.9$ (Yellow) and $\eta = 0.8$ (Green), we see that the $r_2$ values are higher than the $r_1$ values, indicating that a two-cluster state is present in the system, which is the expected outcome. 

A more generalized approach to further this study would be to incorporate pair-wise interactions along with the triadic ones to give a model like:
\begin{equation}\label{MM-HO-M1ext}
\begin{aligned}
    m\ddot{\theta_i} &= \omega_i - \dot{\theta_i} + \frac{K_1}{N} \sum_{j=1}^{N} \sin(\theta_j - \theta_i)\\
    &+ \frac{K_2}{N^2} \sum_{j=1}^{N} \sum_{k=1}^{N} \sin(\theta_j + \theta_k - 2\theta_i)
\end{aligned}
\end{equation} 
which in the mean-field format and rotating frame, with $\psi_1$ set to zero, reads as $m\ddot{\theta_i} = \omega_i - \dot{\theta_i} - K_1 r_1 \sin(\theta_i) - K_2 r_1^2 \sin(2\theta_i)$. However, the simultaneous presence of the first and second-order harmonics of the sinusoidal term makes it complicated to analytically study the parameter space of Eq.~\ref{MM-HO-M1ext}.

\end{document}